\documentstyle[pre,aps,amssymb,psfig,twocolumn,preprint,tighten]{revtex}
\begin{document}
\title{A walk in the parameter space of L--H transitions without
stepping on or through the cracks}
\author{BALL Rowena and DEWAR Robert L.}
\address{Department of Theoretical Physics
and Plasma Research Laboratory\\
Research School of Physical Sciences \& Engineering\\
The Australian National University, Canberra ACT 0200 Australia\\
e-mail: Rowena.Ball@anu.edu.au, Robert.Dewar@anu.edu.au}
\maketitle
\abstract{A mathematically and physically sound
three-degree-of-freedom
dynamical model
that emulates low- to high-confinement mode (L--H) transitions is elicited
from
a singularity theory critique of earlier fragile models.
We construct a smooth map of the parameter
space that is consistent both with the requirements of
singularity theory and with the physics of the process. The model is
found to contain two
codimension 2 organizing centers and two Hopf bifurcations, which underlie
  dynamical behavior that has been observed around L--H transitions but
not mirrored in previous models.
The smooth traversal of parameter space provided by this analysis
gives qualitative guidelines for controlling access to H-mode and
oscillatory r\'{e}gimes. }

\section{Introduction}
A unified, low-dimensional description of the dynamics of
L--H transitions \cite{Connor:2000} would be a valuable aid for the
predictive design and control of confinement states in fusion plasmas.
In this work we report significant progress made toward this goal by
developing the singularity theory approach to modeling
L--H transitions that was introduced in \cite{Ball:2000}. The results give
new insights into the role of energy exchange and dissipation
in the onset, evanescence, and extinction of discontinuous and oscillatory
action in confined plasmas.

The title of this paper refers to the philosophy of singularity
theory \cite{Golubitsky:1985} as applied to dynamical models:
that paths through  parameter space should be
smooth and continuous, and that parameters should be
independent and not fewer than the codimension$^\ast$
~of the system.

Since 1988 \cite{Itoh:1988} many efforts
have been made to derive unified low-dimensional dynamical models that
mimic L--H transitions and/or associated oscillatory behavior
\cite{Shaing:1989,Hinton:1991,Dnestrovskij:1992,Diamond:1994,Carreras:1994b,%
Pogutse:1994,Voj:1995,Sugama:1995,Lebedev:1995,Haas:1995,%
Drake:1996,Hu:1997,Takayama:1998,%
Kardaun:1998,Staebler:1999,Thyagaraja:1999,Odblom:1999}. All of these models
have contributed to the current view in which the
coupled evolution of poloidal shear flow and turbulence
establishes a transport barrier. However, as was shown in \cite{Ball:2000},
the models often
founder at singularities. Consequently, much of the
discussion in the literature concerning the bifurcation 
properties of L--H transition models is qualitatively
wrong.

We examine the bifurcation structure of a semi-empirical
dynamical model for L--H transitions \cite{Diamond:1994},
and find it needs two major operations to give it
mathematical consistency: (1) a degenerate
singularity is identified and unfolded,~(2)~the dynamical state space is
expanded to three dimensions.

We then analyse the bifurcation structure of the
  enhanced model obtained from these operations, the BD model,
  and find it consistent with many known
   features of L--H transitions.
In particular, this
 is the first model that can emulate the onset and {\em abatement} of
  oscillations in H-mode, and direct jumps to oscillatory H-mode
   \cite{Zohm:1995,Shats:1997}.

\section{Bifurcation structure of the DLCT model}
This paradigmatic 2-dimensional model \cite{Diamond:1994} comprises an
evolution equation for the turbulence coupled with
an equation for the flow shear dynamics derived from
the poloidal momentum balance:
\begin{eqnarray}
\frac{dN}{dt}&=&\gamma N-\alpha FN - \beta N^2\label{N'1}\\
\frac{dF}{dt}&=& \alpha FN-\mu F\label{F'1}.
\end{eqnarray}
$N$ is the normalized level of density fluctuations, $F$ is the square of the
averaged {\bf E}$\times${\bf B} poloidal flow shear. The fluctuations grow
linearly with coefficient $\gamma$ and are damped
quadratically with coefficient $\beta$.
The exchange coefficient $\alpha$ is related to
the Reynolds stress, and the damping rate $\mu F$ is due to
viscosity.$^\dagger$ 

\begin{figure}\vspace*{2cm}
\hbox{\hspace*{1cm}
\psfig{file=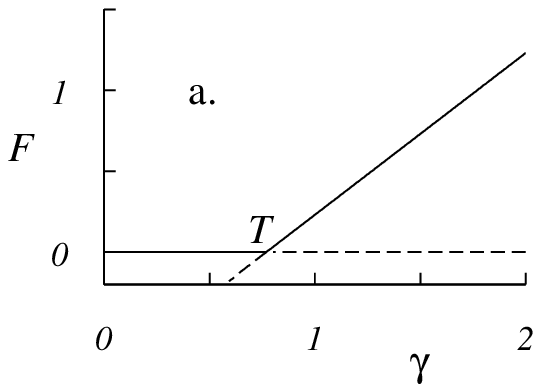}
\psfig{file=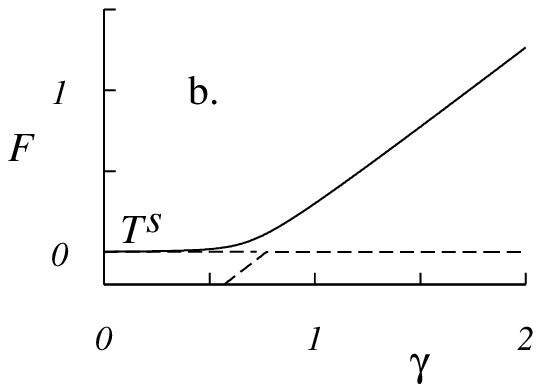}}
\hbox{\hspace*{1cm}
\psfig{file=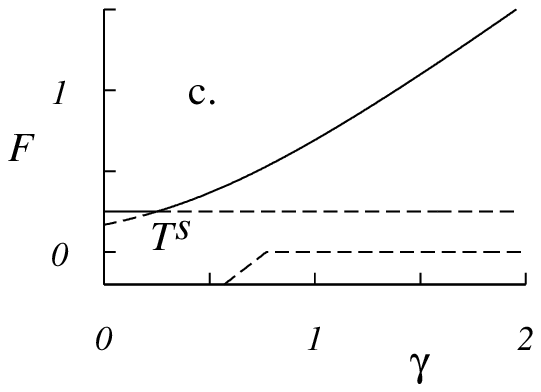}
\psfig{file=fig1d2.ps}}\vspace*{0.5cm}
\caption{\label{fig1}
Bifurcation diagrams of the DLCT model showing how a perturbation of
the shear flow unfolds $T$ but introduces $T^s$.
$\alpha=1$, $\beta=0.77$, $\mu=1$. a. $\varphi=0$, b. $\varphi=0.05$,
c. $\varphi=0.5$, d.~$\varphi=1$.
The region $F < 0$ is not within the phase space but is included to
make the nature of $T$ clearer.
}
\end{figure}

Following the procedure 
outlined in \cite{Ball:2000} we 
form the bifurcation function
$ g = X(F,\gamma)$,
and identify the singular points where $g=g_F=0$.
We find the unique physical singularity
$(F,\gamma)_T=\left(0,\;\beta\mu/\alpha\right)$,
which satisfies the additional defining conditions for a transcritical
bifurcation:
\begin{equation}\label{trans}
g_\gamma=0,\quad g_{FF}\neq 0,\quad \det d^2g<0,
\end{equation}
where $\det d^2g$ is the Hessian matrix of second partial derivatives with
respect to $F$ and~$\gamma$. Evaluating~(\ref{trans}) at $T$
gives $g_\gamma=0$, $g_{FF}=-2\alpha^2/\beta$, $\det d^2g=-\alpha^2/\beta^2$.
  The bifurcation diagram showing the transcritical point $T$
is plotted in Fig. \ref{fig1}a. (In this and subsequent diagrams stable
solutions are indicated by continuous lines and unstable
solutions by dashed lines.)

However, Fig. \ref{fig1}a does not represent the complete bifurcation
structure of the DLCT model because of the following generic
property of the transcritical
bifurcation: {\em it is non-persistent to an arbitrarily small perturbation.}
Since the poloidal shear flow $v^\prime$ is symmetric
under the transformation $v^\prime\rightarrow -v^\prime$ it is appropriate
to introduce the perturbation term  $\varphi F^{1/2}$.
(Note that $F\propto v^\prime{^2}$.) Thus, the modified DLCT model is
\begin{eqnarray}
\frac{dN}{dt}&=&\gamma N-\alpha FN - \beta N^2\eqnum{\ref{N'1}}\\
\frac{dF}{dt}&=& \alpha FN-\mu F + \varphi F^{1/2}\label{F'2}.
\end{eqnarray}

The perturbation term in Eq. \ref{F'2} represents a physically inevitable
  source of shear flow that breaks the symmetry of the internal
shear flow generation and loss rates.
The physics comes from non-ambipolar ion orbit losses
that produce a driving torque for the shear flow.
This contribution to the total shear flow evolution can be quite large
\cite{Kiviniemi:2000},
and in fact the
early models for L--H transitions relied exclusively on a large, nonlinear
ion orbit loss rate
\cite{Itoh:1988,Shaing:1989}. However, ion orbit loss alone cannot
explain turbulence suppression. Here we treat this
term as part of a more complete picture of L--H transition dynamics, and
emphasize its symmetry-breaking nature by assigning the simplest consistent
form to it while recognizing that $\varphi$
  may be a nonlinear function $\varphi(\zeta)$, where
  $\zeta$ may include dynamical variables and parameters.

  Some bifurcation diagrams for increasing values of $\varphi$ are plotted in
  Fig. \ref{fig1}b--d. We see immediately that solution of one problem
  causes another: a nonzero perturbation term does indeed unfold the degenerate
  singularity $T$, but it releases {\em another} degenerate
  singularity~$T^s$.

Before proceeding with a treatment of the new bifurcation $T^s$
we highlight three important issues:

\noindent 1.
Since $\varphi$ is inevitably nonzero in
experiments, no transition can occur at all in the vicinity of $T$, neither
first-order or second-order, contrary to what is stated in \cite{Diamond:1994}.

\begin{figure}\vspace*{2cm}
\hspace*{2cm}\psfig{file=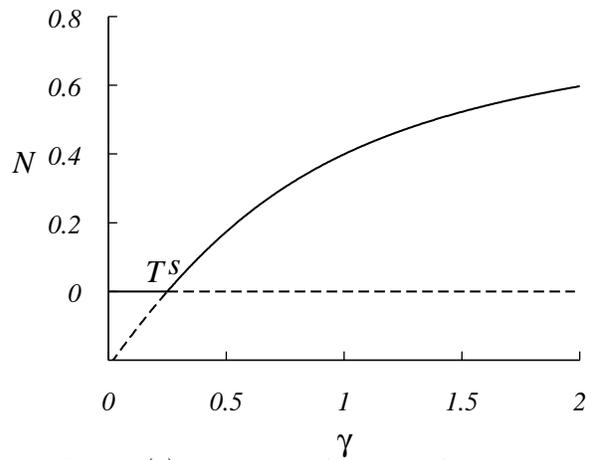}
\caption{\label{fig2} Same as Fig. \ref{fig1}(c) except with $N$
as the state variable.
The region $N < 0$ is not within the phase space but is included to
make the nature of $T^S$ clearer.
}
\end{figure}

\noindent 2.  Both $N$ and $F$ change continuously with $\gamma$ in
the same direction.  The fact that $T$ is a transcritical bifurcation tells us
that only two parameters --- $\varphi$ and any one of the other parameters ---
are required to define the qualitative structure of the problem.
The bifurcation diagram with $N$ as
state variable is plotted in Fig. \ref{fig2}, which should be compared
with Fig. \ref{fig1}c.
As it stands, the model therefore cannot emulate turbulence stabilization
by the shear flow, contrary to what is stated in \cite{Diamond:1994}.

\noindent 3.  To ascertain whether the model can exhibit periodic dynamics
as stated in \cite{Diamond:1994} we
look for a pair of purely complex
conjugate eigenvalues. For Eqs~\ref{N'1} and~\ref{F'2}
  the defining conditions for Hopf bifurcations may be
expressed as
\begin{equation}\label{h}
  g=\text{tr}J=0,\quad \det J > 0,\quad \frac{d}{d\gamma}\text{tr}J\neq 0 ,
  \end{equation}
where $J$ is the Jacobian matrix. We find that $\det J<0$ where
the equalities in Eq. \ref{h} are fulfilled, therefore 
oscillatory dynamics arising from Hopf bifurcations cannot occur. 

This does not rule out the possible existence of periodic
behavior arising from rare and pathological causes. According to Dulac's
criterion \cite{Jordan:1987} Eqs \ref{N'1} and~\ref{F'2} possess {\em no}
periodic solutions arising from {\em any} cause if
there exists $\mathfrak D$ such that
the quantity
\begin{equation}\label{dulac}
S=\frac{\partial}{\partial N}\left({\mathfrak D}W\right) +
   \frac{\partial}{\partial F}\left({\mathfrak D}Y\right)
\end{equation}
never changes sign. Here $W=W(N,F)\equiv dN/dt$, $Y=Y(N,F)\equiv dF/dt$, and
the Dulac function ${\mathfrak D} = {\mathfrak D}(N,F)$ is a real positive
function. Choosing ${\mathfrak D}=1$ we find that
\begin{equation}\nonumber
  S = \alpha\left(N-F\right) - 2N\beta +\gamma -\mu + \varphi F^{-1/2}/2,
  \end{equation}
which clearly can switch sign. However, there may exist a more exotic
Dulac function that forbids a change of the sign of $S$. We have not found
oscillatory solutions numerically in this system.

Returning to the new singularity $T^s$ we find that it is also a
transcritical bifurcation: the conditions (\ref{trans}) evaluated at
$(F,\gamma)_{T^s}=(\varphi^2/\mu^2,\;\alpha\varphi^2/\mu^2)$ yield
$g_\gamma=0$, $g_{FF}=-\beta\mu^6/(2\alpha^2\varphi^4) - \mu^3/\varphi^2$,
$\det d^2g=-\mu^6/(4\alpha^2\varphi^4)$. Does the DLCT model therefore
require a {\em second} perturbation term, this time to
  Eq. \ref{N'1}, to unfold $T^s$?

We remark here that
often there is more than one
universal unfolding for a given bifurcation problem, and we turn to the physics
to decide which is physically consistent.
For the perturbation
in Eq.~\ref{F'2} that unfolded $T$ we chose the form $\varphi F^{1/2}$
because it is physically inevitable that the symmetry
$v^\prime\rightarrow -v^\prime$ be broken.
However, there is no matching physics for a
similar term 
in Eq.~\ref{N'1}.
  Another possibility is that $T^s$ is
spurious, created by an unwarranted collapse of a larger
state space. This idea leads to a suggestion that {\em is} supported by
the physics, that another dynamical variable is intrinsic to a
low-dimensional description of L--H transition dynamics.

\section{Intrinsic 3-dimensional dynamics of L--H transitions}
We introduce the the third dynamical variable by assuming that
$\gamma=\gamma(P)$, where $P$ is the pressure gradient, as have
  a number of other authors
\cite{Carreras:1994b,Haas:1995,Lebedev:1995,Sugama:1995,Thyagaraja:1999}.
Assuming the simplest evolution of $P$ and that $\gamma(P)=\gamma P$,
we arrive at the following
augmented model, obtained purely
  from dynamical and physical considerations:
\begin{eqnarray}
\varepsilon\frac{dP}{dt}&=&q - \gamma P N\label{P'1}\\
\frac{dN}{dt}&=&\gamma P N-\alpha FN - \beta N^2\label{N'2}\\
\frac{dF}{dt}&=& \alpha FN-\mu F + \varphi F^{1/2}.\eqnum{\ref{F'2}}
\end{eqnarray}
In Eq. \ref{P'1} $q$ is the power input and $\varepsilon$ is a
dimensionless parameter that regulates the contribution of the pressure
gradient dynamics to the overall evolution.
The dynamics
is
essentially 3-dimensional with $\varepsilon\approx O(1)$, but
for $\varepsilon\ll 1$ or $\varepsilon\gg 1$ the system can
evolve in two timescales:

\noindent 1.
  The original ``slow'' time $t$.
For $\varepsilon \rightarrow 0$, $\varepsilon dP/dt\approx 0$ and
$P\approx q/(\gamma N)$. The system collapses smoothly to
\begin{eqnarray}
\frac{dN}{dt}& =&q -\alpha FN - \beta N^2\label{N'3}\\
\frac{dF}{dt}& =& \alpha FN-\mu F + \varphi F^{1/2}\eqnum{\ref{F'2}}.
\end{eqnarray}
The organizing center is the unique transcritical bifurcation
$(F,q,\varphi)_T=(0,\;\beta\mu^2/\alpha^2,\;0)$, the spurious $T^s$
is non-existent, and there are no Hopf bifurcations.
For $\varepsilon\gg 1$ we define $\delta\equiv 1/\varepsilon$ and
multiply Eq. \ref{P'1} through by $\delta$; taking the limit as
$\delta \rightarrow 0$ gives $dP/dt\approx 0$, from which $P=P_0$.
We recover the same form as Eqs~\ref{N'1} and~\ref{F'2},
\begin{eqnarray}
\frac{dN}{dt}& =&\gamma P_0 N -\alpha FN - \beta N^2\eqnum{$1^\prime$}\\
\frac{dF}{dt}& =& \alpha FN-\mu F + \varphi F^{1/2}\eqnum{\ref{F'2}},
\end{eqnarray}
along with the ``good'' bifurcation $T$ and the ``bad'' bifurcation $T^s$
--- therefore we suggest that this is a non-physical limit for
$\varepsilon$.

\noindent 2.  In ``fast'' time $\tau \equiv \varepsilon/t$ and,
recasting the system accordingly,
it can be seen that on this timescale the dynamics
becomes
1-dimensional in
$P$ in both limits.

The
organizing
center of the bifurcation problem obtained from
Eqs \ref{P'1}, \ref{N'2}, and \ref{F'2} is the unique transcritical
bifurcation
$(F,q,\varphi)_T=(0,\;\beta\mu^2/\alpha^2,\;0)$,  $g_{FF}=-\alpha^2/\beta$,
$\det d^2g = -\alpha^4/(4\beta^2 \mu^2)$, and the spurious singularity
$T^s$ is non-existent. We now have the bones of an improved dynamical
model for L--H transitions, but it still does not
emulate the following characteristics of L--H
transitions:
(a)~Hysteresis: Since there is no non-trivial point where $g_{FF}= 0$ it cannot
model discontinuous transitions or classical hysteretic
behavior. (b)~Oscillations in H-mode: These have not been found numerically.
In a 3~-~dimensional dynamical system it is, of course, very
difficult to prove that oscillatory solutions do {\em not} exist.

Evidently we need {\em more} nonlinearity or {\em higher order} nonlinearity
to produce enough
competitive
interaction.
  To obtain multiple solutions,
at least,  the bifurcation equation $g$
should map to the normal form for the pitchfork bifurcation
$h=\pm x^3\pm \lambda x$.

Several authors have taken the viscosity coefficient as
a function of the pressure gradient,
but usually it is treated as a constant. In \cite{Sugama:1995} the viscosity
was considered to be the sum 
of neoclassical and anomalous or turbulent contributions,
both with separate power-law
dependences on the pressure gradient. We shall adopt this bipartite form
and in Eq. \ref{F'2} take
\begin{equation}\label{vis}
\mu=\mu(P)= \mu_{\text{neo}}P^n+\mu_{\text{an}}P^m.
\end{equation}
Equations \ref{P'1}, \ref{N'2} and \ref{F'2} with (\ref{vis}) comprise
the BD model.

The values of the exponents $n$ and $m$ are not precisely known  empirically or
from theory. In
\cite{Itoh:1999} $\mu_{\text{an}}$ is given as having
a $P^{3/2}$ dependence, but is also subject to the additional
influence of a $P$-dependent curvature factor.
In this work we take
$n=-3/2$ as in \cite{Sugama:1995} and $m=5/2$.

\section{Bifurcation structure of the BD model}
The bifurcation problem obtained from the BD model
  contains two codimension 2 organizing centers:

\noindent 1. The defining conditions for the pitchfork,
\begin{equation}\label{pf}
g=g_F=g_{FF}=g_q=0,\; g_{FFF}\neq 0,\; g_{Fq}\neq 0,
\end{equation}
find this singularity occurring at $(F,q,\beta,\varphi)_\wp=(0,\;
8\mu_{\text{an}}^{1/8}\mu_{\text{neo}}^{7/8}\gamma/(7^{7/8}\alpha),\;
(7^{3/8}\alpha\gamma)/(8\mu_{\text{an}}^{5/8}\mu_{\text{neo}}^{3/8}),\;0)$,
$g_{FFF}=-12(7^{7/8})\mu_{\text{an}}^{7/8}\mu_{\text{neo}}^{1/8}\gamma/\alpha$,
$g_{Fq}=2(7\mu_{\text{an}}/\mu_{\text{neo}})^{1/4}$.
The pitchfork $\wp$ becomes a transcritical bifurcation $T^l$ away from the
critical value of $\beta$.

\noindent 2. Another transcritical bifurcation $T^u$ occurs at
  $(F,q,\varphi)_{T^u}
=(0,\,P^2\gamma/\beta,\,0)$,
$g_{FF}=-2P\gamma^3(7P^{5/2}\alpha\gamma-8\mu_{\text{neo}}\beta)/
(P^{7/2}\alpha\beta\gamma^2)$,
$\det 
d^2g=-(-3P^{5/2}\alpha\gamma+8\mu_{\text{neo}}\beta)^2/(4P^7\alpha^2\gamma^2)$. 

$T^l$ and $T^u$ are annihilated at a second
codimension 2 bifurcation. The defining conditions for this point are
\begin{equation}\label{c22}
g=g_F=g_q=\det d^2g=0,\; g_{FF}\neq 0,\; g_{Fq}\neq 0
\end{equation}
  At this point we find\linebreak $(F,q,\beta,\varphi)=
(0,(8(7^{1/8})\mu_{\text{neo}}\gamma)/
(3\alpha(\mu_{\text{neo}}/\mu_{\text{an}})^{1/8}),$\linebreak$
3(\mu_{\text{neo}}/\mu_{\text{an}})^{5/8}\alpha\gamma/(8 
(7^{5/8}\mu_{\text{neo}})),0),$\linebreak
$g_{FF}= -64 (7\mu_{\text{an}})^{5/8}\mu_{\text{neo}}^{3/8}\gamma/(3\alpha)$,
$g_{Fq}=4(7\mu_{\text{an}}/\mu_{\text{neo}})^{1/4}$.

\begin{figure}\vspace*{2cm}
\hspace*{2cm}\vbox{\psfig{file=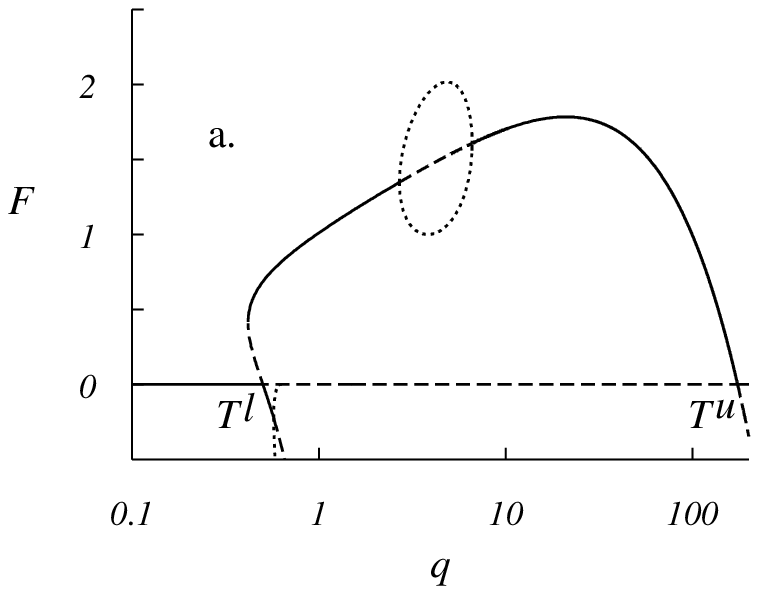}
\psfig{file=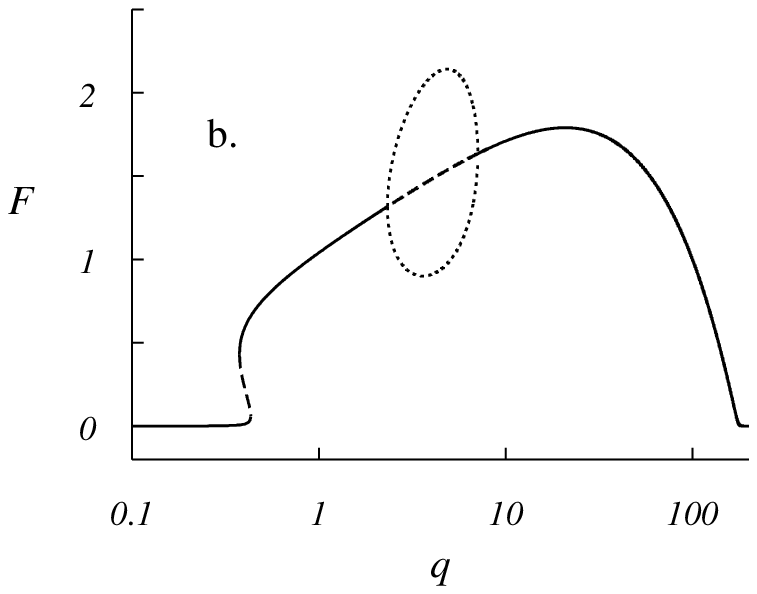}}\vspace*{5mm}
\caption{\label{fig3} Bifurcation diagrams BD model.
$\alpha=2.4$, $\beta=1$, $\gamma=1$,
$\varepsilon=1$, $\mu_{\text{neo}}=1$, 
$\mu_{\text{an}}=0.05$,$n=-1.5$, $m=2.5$.
a. Bifurcation
structure of the partially perturbed system, with $\varphi=0$.
b. $\varphi=0.05$.
For clarity the lower unstable branches are not plotted in (b). }
\end{figure}

In Fig \ref{fig3}a the partially perturbed bifurcation diagram is plotted,
showing the lower and upper transcritical bifurcations $T^l$ and $T^u$.
In Fig. \ref{fig3}b the fully  perturbed, physical  bifurcation diagram
is plotted, where $\varphi>0$.
There are also {\em two} Hopf bifurcations on the upper H-mode branch
in Fig. \ref{fig3}
  linked by a branch of stable limit cycles.
  The dotted lines mark the maximum and minimum
  amplitude trace of the limit-cycle branch.
  This reflects the passage through an oscillatory r\'{e}gime
   that is often observed in experiments.

\begin{figure}\vspace*{2cm}
\hspace*{2cm}\psfig{file=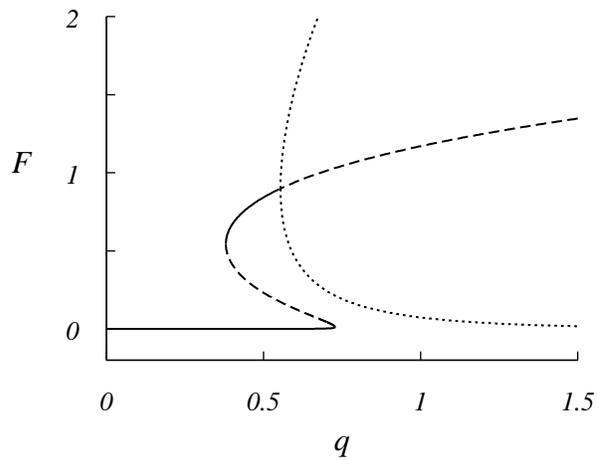}\vspace*{5mm}
\caption{\label{fig5} Under extreme conditions, where the turbulence 
dissipation
rate is low relative to the generation rate, the jump at the
lower limit point can occur directly to an oscillatory state on the
H-mode branch. $\beta=0.1$, other parameters as for Fig. \ref{fig3}.}
\end{figure}

Since it is a codimension 2 bifurcation problem, the qualitative
structure is fully defined by $q$
and two auxiliary parameters. One of these is obviously $\varphi$, the other
may be any one of the other parameters. We choose $\beta$ because
we are interested in the 
effects of poor turbulence dissipation (i.e. low $\beta$).
Figure \ref{fig5} illustrates how a jump can occur directly to oscillatory 
states, a phenomenon which is frequently observed.

\begin{figure}\vspace*{2cm}
\hspace*{2cm}\psfig{file=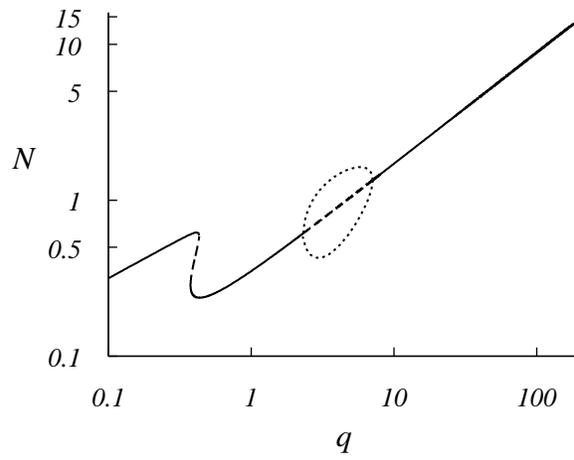}
\caption{\label{fig6} Same as Fig. \ref{fig3}, except with $N$ as state
variable.}
\end{figure}

Figure \ref{fig6}, to be compared with Fig. \ref{fig3}b, shows
that the BD model does indeed reflect shear flow
suppression
of turbulence.

\section{Discussion and conclusions}
A dynamical model that emulates much of the typical behavior
around L--H transitions has been elicited from an earlier fragile
model that had serious flaws by considering the relationship between
bifurcation structure and the physics of the process.
Built in to this model are the following major dynamical features
of L--H transitions:

\noindent 1. Discontinuous, hysteretic transitions, or smooth changes 
with power
input, depending
on the degree of turbulence dissipation $\beta$, or equivalently,
the viscosity.

\noindent 2. Two Hopf bifurcations in H-mode. It is the first model that can
emulate the onset {\em and} abatement of oscillatory behavior, and
a transition directly into oscillatory H-mode.

\noindent 3. Turbulence
suppression
by the shear flow.

\noindent 4. A maximum in the shear flow generated by the turbulence, 
followed by
a decrease as the power input flowing to the turbulence is raised.

\noindent 5. Turbulence generation from non-ambipolar losses.


Finally, we note that the existence of {\em two} codimension~2 bifurcations
  is suggestive: Should there be 
an expansion of the system,  
  perhaps expressing fluctuations of the magnetic field, that creates
  (or annihilates) the two bifurcations at a codimension 3 singularity?
  In other words, does a more complete model contain an
  organizing center of higher order? In singularity theory we
  persevere in seeking higher order behavior: that is how the relationship
  between a model and the process it represents is tracked.
  This question is currently under investigation.
  

\begin{thebibliography}{10}
\bibitem[*]{endnote-1}
Codimension is defined as the minimum number of independent auxiliary
   parameters required for a universal unfolding of the most degenerate
   singularity, or organizing center, in the system. A universal unfolding
   captures all possible qualitative bifurcation states of a given bifurcation
   problem.
\bibitem[\dagger]{endnote-2}
To avoid confusion in later discussion some minor changes to the original
   notation have been made. Also, for convenience a factor of 2 is absorbed into
   the coefficients, that arises because the dynamical variables are squared
   quantities,
   implying that the phase space is restricted to the first quadrant.
\end{thebibliography}

Acknowledgment: This work is supported by an
Australian Research Council Postdoctoral Fellowship.

\onecolumn

\clearpage

\clearpage

\clearpage

\clearpage

\end{document}